\documentstyle[prl,aps,twocolumn]{revtex}
\begin{document}
\preprint{Nov. 1995}
\draft
\title{Magnetic Structure of Rare-Earth Nickel Boride-Carbides}
\author{V. A. Kalatsky}
\address{\em Department of Physics, Texas A\&M University,
College Station, Texas 77843-4242}
\author{V. L. Pokrovsky}
\address{\em Department of Physics, Texas A\&M University,
College Station, Texas 77843-4242  \\
and Landau Institute for Theoretical Physics, Kosygin str.2,
Moscow 117940, Russia}
\date{November 28, 1995}
\maketitle
\begin{abstract}
We present a theory of the magnetic rare-earth nickel boride-carbides
($\mbox{RENi}_2\mbox{B}_2\mbox{C}$, RE=Ho, Er, Tm). Large total
angular momentum allows one to employ semiclassical approximation and
reduce the problem to the four-positional clock model. The latter gives
a proper phase diagram in magnetic field-magnetic moment plane for
varying directions of in-plane magnetic field.
The theory explains recent experimental observations.
\end{abstract}

A complex phase diagram of $\mbox{HoNi}_2\mbox{B}_2\mbox{C}$
in the tem\-pe\-ra\-tu\-re--mag\-ne\-tic field plane
has been discovered in the neutron scattering experiments~\cite{holsingle}
and in measurements of resistivity~\cite{Don}.
Recent measurements of the anisotropic magnetization of
$\mbox{HoNi}_2\mbox{B}_2\mbox{C}$
by P.~Canfield and coworkers~\cite{can1,can2}
revealed a complicated magnetization--magnetic field diagram:
re-entrant behavior for the low fields and a set of magnetic
transitions for the high fields. The magnetic moment is highly anisotropic
and has a strong dependence upon direction of magnetic field
in the basal $a$-$b$ plane.
A purpose of this article is to present a simple model for
the magnetic subsystem in $\mbox{RENi}_2\mbox{B}_2\mbox{C}$,
explaining the experimental facts.

The crystal structure of the $\mbox{RENi}_2\mbox{B}_2\mbox{C}$
has been studied by
T.~Siegrist {\em et al.}~\cite{Lustructure} (RE=Lu) and
Q.~Huang {\em et al.}~\cite{Hostructure} (RE=Ho).
The $\mbox{RENi}_2\mbox{B}_2\mbox{C}$ compounds have
the body-centered-tetragonal (bct) structure (space group $I4/mmm$).
A simplest form of the single-ion Hamiltonian compatible with
the crystal field symmetry is
\begin{equation}
{\cal H}_{CEF}=\frac{a}{2}J_z^2+\frac{b}{2}(J_{+}^4+J_{-}^4),
\label{cef2}
\end{equation}
where $J_{\pm}$, $J_z$ are components of the total angular momentum,
$a$ and $b$ are crystal field constants.
We assume that moments are presumably aligned in the $a$-$b$ plane
and neglect terms proportional to $J_z^4$ in comparison to those
with $J_z^2$.
The constant $a$ must be
positive to fix {\bf J} in the $a$-$b$ plane.

The fourth-order terms in Eq.~(\ref{cef2}) lift the $O(2)$ degeneracy
in favor of the four-fold symmetry.
The Hamiltonian~(\ref{cef2}) can be treated quasi-classically
since $J_x$ and $J_y$ are rather large (J=8, 15/2, and 6
for Ho, Er, and Tm, respectively).
We introduce the angle $\phi$ for the orientation of the
moment {\bf J} in the $a$-$b$ plane, so that $J_x=J\cos\phi$ and
$J_y=J\sin\phi$.
Then Hamiltonian~(\ref{cef2}) can be presented in the form
\begin{equation}
{\cal H}=-\frac{a}{2}\frac{\partial^2}{\partial\phi^2}
+bJ^4(1+\cos(4\phi))-h\cos(\phi-\phi_h),
\label{ham}
\end{equation}
where $h$ is the absolute value of the external magnetic field multiplied
by $5J\mu_B/4$, $\phi_h$ is the angle between direction of
magnetic field and reference tetragonal axis ({\em e.g.} $a$ axis).
Exploiting again a large value of $J$, it is natural to assume that
$2bJ^4\gg a$. Then the potential energy
in (\ref{ham}) has deep wells at directions
$\phi=\pm\pi/4,\quad\pm3\pi/4$.
Non-zero magnetic field destroys the tetragonal symmetry,
but the symmetry breaking is small unless $h$ exceeds
the value $bJ^4$ corresponding to fields essentially large than
20~T~\cite{foot}.
Thus, initially continuous moment {\bf J} is reduced to a discrete variable
taking only four values. This is a kind of so-called clock model with
4 positions of the ``hand".

Let us denote 4 states of the ``hand" as $|k\rangle$, where $k=0,\ldots,3$.
Then the Hamiltonian (\ref{ham}) is equivalent to $4\times4$ matrix

\begin{equation}
\hat{{\cal H}}_{kk'}=\epsilon_k\delta_{kk'}+
w(\delta_{k,k'-1}+\delta_{k,k'+1}),
\label{matrix}
\end{equation}
where $\epsilon_k=-h\cos((2k+1)\pi/4-\phi_h)$ and $w$ is
a matrix element of the ``kinetic energy"
$\frac{a}{2}\frac{\partial^2}{\partial\phi^2}$
between adjacent states of the ``hand".
In the approximation $h\ll bJ^4$ all the overlap integrals are
the same ($w$).
The diagonalization of the matrix (\ref{matrix}) can be
performed explicitly at any values of the parameters $h$, $\phi_h$, and $w$.
The lowest eigenvalue gives the energy of the ground state
\begin{equation}
E=-\sqrt{\frac12\left(4w^2+h^2+\sqrt{(4w^2+h^2)^2-h^4
\cos(2\phi_h)^2}\right)}.
\label{oneionenergy}
\end{equation}
The result~(\ref{oneionenergy}) explains the orientation dependent
saturation in high magnetic field~\cite{can1,can2}
(see Fig.~\ref{fig1}).
Namely, it has been observed in~\cite{can2} that the magnetization
along magnetic field reaches saturation
values depending on direction according to the law
$M_s(\phi_h)\propto\cos(\pi/4-\phi_h)$. The moments in such a state
are directed along an easy direction closest to the direction
of the field. The direction-dependent saturation
Fig.~\ref{fig1}
is an intermediate asymptotic corresponding to the field interval
$w\ll h\ll bJ^4$. At higher fields $h\gg bJ^4$ the direction
independent saturation is reached.

Further we employ this simplified description of the
localized moments with minor modification to incorporate the collective
effects. Let us introduce variables $k_r=0,\ldots,3$ at each site
of the lattice and the interaction Hamiltonian
\begin{equation}
{\cal H}_i=\!\sum_{\bf r,r'}K_{\bf r,r'}\cos\frac{\pi}{2}(k_{\bf r}
-k_{\bf r'})+\!\sum_{\bf r,r'}L_{\bf r,r'}\cos\pi(k_{\bf r}-k_{\bf r'}),
\label{interaction}
\end{equation}
where $K_{\bf r,r'}$ and $L_{\bf r,r'}$ are interaction energies.
Only nearest neighbor interaction will be taken into account, so that
the only non-zero constants are
$K_{\bf r,r\pm a}=K_{\bf r,r\pm b}=-\frac12K_{\parallel}$,
$L_{\bf r,r\pm a}=L_{\bf r,r\pm b}=-\frac12L_{\parallel}$,
$K_{{\bf r,r\pm c/}2}=K_{\perp}$ and
$L_{{\bf r,r\pm c/}2}=L_{\perp}$,
where ${\bf a, b, c}$ are the primitive
vectors of the lattice. We assume that
$K_{\parallel}>K_{\perp}>0$,
$|L_{\parallel}|>|L_{\perp}|$.
For simplification we put
($L_{\parallel}=L_{\perp}=0$).

In this article we find only the ground state of the total
Hamiltonian ${\cal H}={\cal H}_s+{\cal H}_i$ where
the single-ion Hamiltonian ${\cal H}_s$ is the direct sum
of matrices~(\ref{matrix}). The ground state will be supposed
to be homogeneous in-plane.
The existing experimental data have no agreement upon the in
plane structure of $\mbox{HoNi}_2\mbox{B}_2\mbox{C}$.
The experiments performed on single
crystals by Goldman, {\em at al.}~\cite{holsingle},
have clear presence of the satellites
at $0.585 {\bf a}^{\ast}$ in the temperature range of 5--6 K.
On the other hand neutron powder diffraction
measurements by Grigereit, {\em at al.}~\cite{holpowder,Hostructure},
did not find evidence of these satellites.
Both groups agree that the structure in the $a$-$b$
plane transits to the ferromagnetically aligned sheets below 5 K.
We will assume ferromagnetic ordering in the $a$-$b$ plane.
This implies that $K_{\parallel}>0$. The antiferromagnetic order in
$c$-direction implies that $K_{\perp}>0$.

We apply a variational
procedure, assuming the total wave function to be a product of
single-ion wave functions.
The wave functions at sites belonging to the same $a$-$b$ plane
are assumed to be identical. Neighboring planes are assumed to
belong to different sublattices 1 and 2.
The single-ion states are described by
two set of variational parameters $\alpha_k, \beta_k$,
$k=0,\ldots,3$ where $\alpha_k$ and $\beta_k$ are the amplitudes for a
``hand" to be in the $k$'th potential well for the first and second
sublattice respectively.
With this premises the energy per site is
\begin{eqnarray}
&&{\cal H}_{var}=
\frac{K_{\perp}}{2}[(\alpha_0^2-\alpha_2^2)(\beta_0^2-\beta_2^2)+
(\alpha_1^2-\alpha_3^2)(\beta_1^2-\beta_3^2)]  \nonumber \\
&&-\frac{K_{\parallel}}{4}[(\alpha_0^2-\alpha_2^2)^2
\!+(\alpha_1^2-\alpha_3^2)^2\!+(\beta_0^2-\beta_2^2)^2
\!+(\beta_1^2-\beta_3^2)^2
] \nonumber \\
&&+w[(\alpha_0+\alpha_2)(\alpha_1+\alpha_3)+
(\beta_0+\beta_2)(\beta_1+\beta_3)]  \nonumber \\
&&-\frac{h_x}{2}[(\alpha_0^2-\alpha_2^2)+
(\beta_0^2-\beta_2^2)]\nonumber\\&&-
\frac{h_y}{2}[(\alpha_1^2-\alpha_3^2)+(\beta_1^2-\beta_3^2)].
\label{varham}
\end{eqnarray}
where $h_x=h\cos(\pi/4-\phi_h)$ and $h_y=h\sin(\pi/4-\phi_h)$,
due to the symmetry the consideration is constricted to the range
$0\leq \phi_h\leq\pi/4$.
The minimization of the Hamiltonian~(\ref{varham})
with the constraints $\sum\alpha_i^2=\sum\beta_i^2=1$
leads to a system of nonlinear equations
which allows a partial separation of variables.
Proper variables are determined by following relations
\begin{eqnarray}
\cos\zeta_1=\alpha_0^2-\alpha_1^2-\alpha_2^2+\alpha_3^2,\nonumber \\
\cos\zeta_2=\beta_0^2-\beta_1^2-\beta_2^2+\beta_3^2, \nonumber \\
\cos\eta_1=\alpha_0^2+\alpha_1^2-\alpha_2^2-\alpha_3^2,\nonumber \\
\cos\eta_2=\beta_0^2+\beta_1^2-\beta_2^2-\beta_3^2.
\label{variables}
\end{eqnarray}
In terms of $\zeta_i$ and $\eta_i$ the minimization conditions read
as follows
\begin{eqnarray}
K_{\perp}\cos\zeta_1-K_{\parallel}\cos\zeta_2-2\cot\zeta_2=h_x-h_y,\nonumber\\
K_{\perp}\cos\zeta_2-K_{\parallel}\cos\zeta_1-2\cot\zeta_1=h_x-h_y,\nonumber\\
K_{\perp}\cos\eta_1-K_{\parallel}\cos\eta_2-2\cot\eta_2=h_x+h_y,\nonumber\\
K_{\perp}\cos\eta_2-K_{\parallel}\cos\eta_1-2\cot\eta_1=h_x+h_y,
\label{system2}
\end{eqnarray}
where $w$ has been set to 1.
The systems of equations~(\ref{system2}) have been solved numerically.

In the general case $0<\phi_h<\pi/4$ there are three phases
and two first order phase transitions
at fields $h_{c1}(\phi_h)$ and $h_{c2}(\phi_h)$.
In the range of small magnetic field the most energy favorable
is antiferromagnetic phase (AF), in which spins in one sublattice are
parallel whereas spins of the neighboring sublattices are antiparallel.
At increasing field it is replaced by the ferrimagnetic phase (FM),
in which spins of the neighboring sublattices are perpendicular.
Finally the spin-flip proceeds into a phase with almost parallel
spins in the neighboring sublattices
(paramagnetic or spin-flip phase (PM)).
In special cases of $\phi_h=0$ and $\phi_h=\pi/4$ the numbers of phases
and phase transitions are reduced by one.
Namely, for $\phi_h=0$ the system undergoes
a transition from the AF into the FM, final
paramagnetic saturation happens for experimentally unreachable
magnetic field $h>bJ^4$.
For $\phi_h=\pi/4$ the system takes up from
the AF directly into the PM.
Numerically found diagram in $h$-$\phi_h$ plane is presented on
Fig.~\ref{fig2} for $K_{\perp}=3$, $K_{\parallel}=4$.

For the two phase transitions and magnetization discontinuities at
the transition points we find
\begin{equation}
h_{c1}=\frac{h_c}{\cos\phi_h},\quad
h_{c2}=\frac{h_c}{\sin\phi_h},
\label{field}
\end{equation}
\begin{equation}
\Delta M_1=M_c\cos\phi_h,\quad \Delta M_2=M_c\sin\phi_h
\label{magnetization}
\end{equation}
where $h_c=h_c(K_{\perp},K_{\parallel})$ and
$M_c=M_c(K_{\perp},K_{\parallel})$. The graphs of $h_c$ and $M_c$ {\em vs.}
$K_{\parallel}$ for some particular values of $K_{\perp}$ are given at
Figs.~\ref{fig3} and~\ref{fig4}, respectively.
The graphs of magnetization {\em vs.} magnetic
field for different direction of the field are shown in
Fig.~\ref{fig5}.

A series of simple relationships can be obtained in a limiting
case of $K_{\parallel}\gg w$, well approaching the conditions of the
experiments.
Then the ``hands" are tightly
bound to the tetragonal easy directions. Magnetic field lifts
four-fold symmetry and one of the ``hands"
(labeled by $\alpha_i$ for definiteness)
is set to the easy direction closest to the direction of magnetic field
$\alpha_0^2=1$ and $\alpha_i=0$, $i=1,2,3$. The other ``hand" can take
three positions ($\beta_3=0$) depending upon strength
and direction of magnetic field.
The three positions correspond to the three phases, {\em i.e.},
AF $\beta_2^2=1$, $\beta_i=0$, $i=0,1$,
FM $\beta_1^2=1$, $\beta_i=0$, $i=0,2$, and
PM $\beta_0^2=1$, $\beta_i=0$, $i=1,2$.
The energies, magnetization and transition lines can be found
straightforwardly resulting in
\begin{equation}
h_c(K_{\perp},K_{\parallel}\rightarrow\infty)=\frac{K_{\perp}}{\sqrt{2}},\quad
M_c(K_{\perp},K_{\parallel}\rightarrow\infty)=\frac{1}{\sqrt{2}}.
\end{equation}
These facts are confirmed clearly by numerical analysis
(see Figs.~\ref{fig3} and~\ref{fig4}).

The results fit reasonably well the experimental facts.
P.~C.~Canfield {\em et. al.}~\cite{can2} have observed
three phase transitions instead of two predicted by our theory.
We denote the experimental lines as $H_{c1}(\phi_h)$,
$H_{c2}(\phi_h)$, and $H_{c3}(\phi_h)$.
It has been found~\cite{can2} that the dependencies are
\[
H_{c1}=\frac{3.36\mbox{kG}}{\cos(\pi/4-\phi_h)},\quad
H_{c3}=\frac{9\mbox{kG}}{\sin\phi_h}.
\]
There has been no evaluation given for $H_{c2}$.
Our critical lines $h_{c1}$ and $h_{c2}$ plausibly
correspond to the experimental
lines $H_{c2}$ and $H_{c3}$ respectively. There is a perfect fit
between $h_{c2}$ (see Eq.~(\ref{field})) and $H_{c3}$
giving $h_{c}\approx9\mbox{kG}$.
We have already noted the excellent agreement between the theory
and experiment for the angular dependence of the saturation magnetization.

The drawback of the model is that one extra phase, observed in the
experiment between lines $H_{c1}$ and $H_{c2}$, is absent in our model.
We presume that this is the chiral phase seen
in the zero-field neutron diffraction experiments~
\cite{holpowder,holsingle} (the satellite $0.915{\bf c}^{\ast}$).
Of course the spiral period might change in the presence
of magnetic field. The spiral phase cannot be principally obtained in the
framework of the two sublattice model.
The number of layers should be at least
equal to the ratio of the spiral structure period to the spacing between
nearest holmium layers. The physical reason for appearance of the
long-periodic modulated phase can be either RKKY forces or quantum
tunneling ($w$) which becomes active when the main collective
interaction $K_{\perp}$ favoring antiferromagnetism is suppressed by
the magnetic field.
Independent estimates of the quantum parameter $w$~\cite{can1} show
that it is rather small ($w\sim$2--2.5~K). Nevertheless, it cannot be
neglected since it determines the magnetic susceptibility in the
low--temperature limit.

We have shown that a simple 4-position clock model stemming from
the quasiclassical single-ion Hamiltonian explains the angular
dependencies of the saturation magnetization and at least one of
the transition points observed in the experiment. We believe that
this fact indicates convincingly that the 4-position model is valid
for the magnetic subsystem of RE nickel boride-carbides.

Incorporating the simplest inter-spin interaction we have found two
phase transitions between AF, FM and spin-flip phases.
We expect that in a slightly modified model modulated phases are the
ground-states in some range of parameters.

Among the predictions which can be checked experimentally we
emphasize following ones: \\
i) In FM phase the magnetization vectors of the
sublattices are perpendicular each other. \\
ii) The jumps of magnetization are simple functions of direction
(see Eq.~(\ref{magnetization})).

Our next purpose is the thermodynamics of the model
as well as consideration of the long period chiral structures.

We are grateful to P.~C.~Canfield and collaborators for the
opportunity to read their articles~\cite{Don,can1} prior
publication and especially to D.~Naugle for numerous discussions.

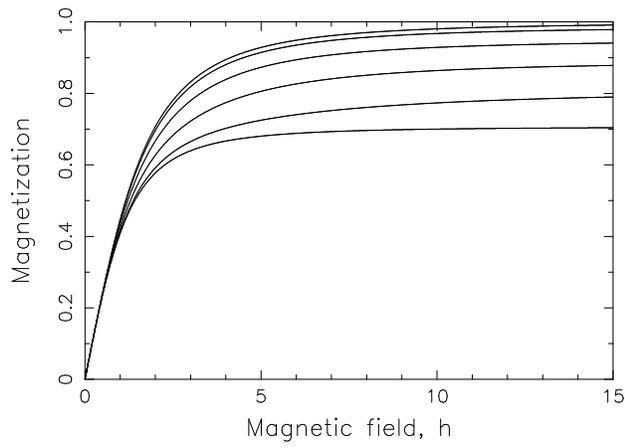
\begin{figure}
\caption{
Magnetization {\em vs.} magnetic field for magnetic field direction
(from top to bottom) $\phi_h=45,\,\,36,\,\,27,\,\,18,\,\,9$, and $0$
degrees, respectively, the single-ion model
(see Eqs.~(\protect\ref{matrix},\protect\ref{oneionenergy})).
$\phi_h=0$ corresponds to the direction $[100]$
and $\phi_h=\pi/4$ to $[110]$.
}\label{fig1}
\end{figure}
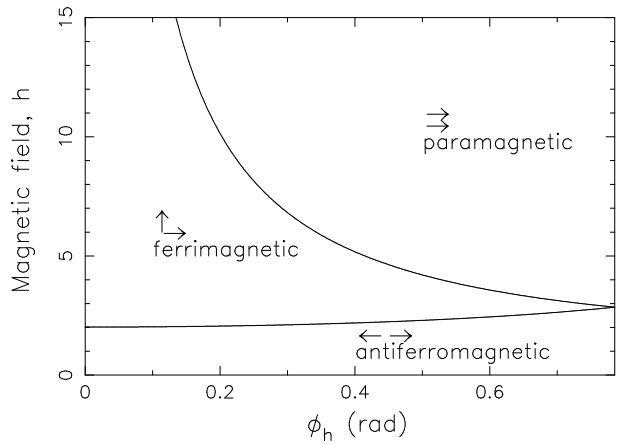
\begin{figure}
\caption{
$h$-$\phi_h$ phase diagram, $K_{\perp}=3$, $K_{\parallel}=4$,
the two sublattice model.
}\label{fig2}
\end{figure}
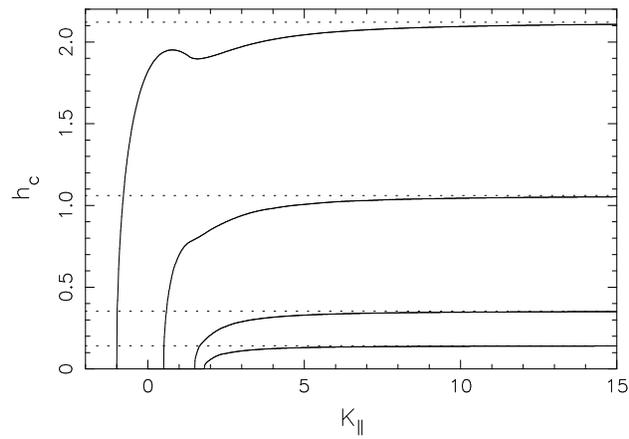
\begin{figure}
\caption{
$h_c$ (see Eq.(\protect\ref{field})) {\em vs.} $K_{\parallel}$,
for interplane interaction (from top to bottom)
$K_{\perp}=3,\,\,1.5,\,\,0.5$, and $0.2$, respectively,
the two sublattice model.
}\label{fig3}
\end{figure}
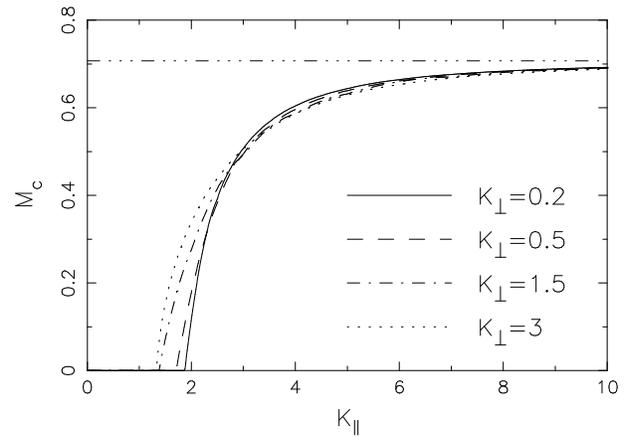
\begin{figure}
\caption{
$M_c$ (see Eq.(\protect\ref{magnetization})) {\em vs.}
$K_{\parallel}$, the two sublattice model.
}\label{fig4}
\end{figure}
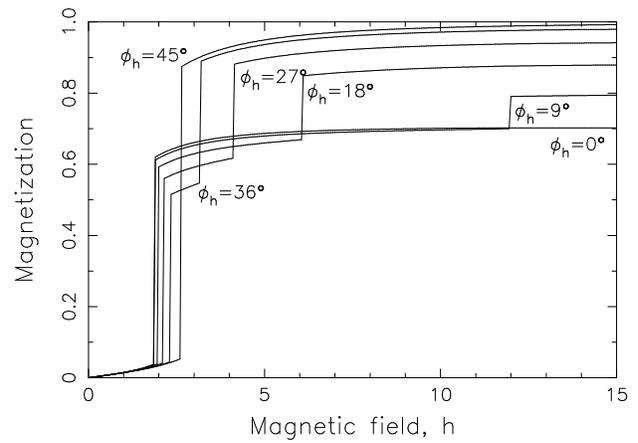
\begin{figure}
\caption{
Magnetization {\em vs.} magnetic field, $K_{\perp}=3$, $K_{\parallel}=4$,
the two sublattice model.
}\label{fig5}
\end{figure}

\begin{references}
\bibitem{holsingle} A.~I.~Goldman, C.~Stassis, P.~C.~Canfield,
J.~Zaretsky, P.~Dervenagas, B.~K.~Cho, D.~C.~Johnston, and
B.~Sternlieb,
Phys. Rev. B {\bf 50}, 9668 (1994).
\bibitem{Don} K.~D.~D.~Rathnayaka, D.~G.~Naugle, B.~K.~Cho, and
P.~C.~Canfield,
Phys. Rev. B in press.
\bibitem{can1} B.~K.~Cho, B.~N.Harmon, D.~C.Johnston, and P.~C.~Canfield,
submitted to Phys. Rev. B.
\bibitem{can2} P.~C.~Canfield,
private communications.
\bibitem{Lustructure} T.~Siegrist, H.~W.~Zandbergen, R.~J.~Cava,
J.~J.~Krajewski, and W.~F.~Peck, Jr.,
Nature {\bf 367}, 254 (1994).
\bibitem{Hostructure} Q.~Huang, A.~Santoro, T.~E.~Grigereit,
J.~W.~Lynn, R.~J.~Cava, J.~J.~Krajewski, and W.~F.~Peck, Jr.,
Phys. Rev. B {\bf 51}, 3701 (1995).
\bibitem{foot} This fact follows from the measurements of saturated
magnetization~\cite{can2} which do not display any tendency to
isotropisation at field $H\sim20$~T.
\bibitem{holpowder} T.~E.~Grigereit, J.~W.~Lynn, Q.~Huang,
A.~Santoro, R.~J.~Cava, J.~J.~Krajevski, and W.~E.~Peck, Jr.,
Phys. Rev. Lett. {\bf 73}, 2756 (1994).
\end{references}
\end{document}